\documentstyle{article}

\newcommand{\non}{non-degenerate closed odd 2-form}
\newcommand{\Lag}{Lagrangian submanifold}
\newcommand{\dar}{Darboux coordinates}
\title{Geometry of Batalin-Vilkovisky quantization.}
\author{Albert Schwarz\\
Department of Mathematics, University of California,\\ Davis, CA 95616\\
ASSCHWARZ@UCDAVIS.EDU}
\begin {document}
 \maketitle
 \smallskip
 \begin{abstract}

 A very general and powerful approach to quantization of gauge theories was
proposed by Batalin and Vilkovisky [1],[2]. The present paper is devoted to the
study of geometry of this quantization procedure. The main mathematical objects
under consideration are $P$-manifolds and $SP$-manifolds (supermanifolds
provided with an odd symplectic structure and, in the case of $SP$-manifolds,
with a volume element). The Batalin-Vilkovisky procedure leads to consideration
of integrals of the form $\int_L Hd\lambda$ where $L$ is a \Lag\ of an
$SP$-manifold $M$ and $H$ satisfies the equation $\Delta H=0$ where $\Delta$ is
an odd analog of Laplacian. The choice of $L$ can be interpreted as a choice of
gauge condition; Batalin and Vilkovisky proved that in some sense their
procedure is gauge independent. Namely they proved that
$\int_{L_0}Hd\lambda_0=\int_{L_1}Hd\lambda_1$ if \Lag s $L_0$ and $L_1$ are
connected by a continuous family $L_t$ of \Lag s. We will prove that the same
conclusion can be made in the much more general case when the bodies $m(L_0)$
and $m(L_1)$ of submanifolds $L_0$ and $L_1$ are homologous in the body $m(M)$
of $M$. This theorem leads to a conjecture that one can modify the quantization
procedure in such a way as to avoid the use of the notion of \Lag . In the next
paper we will show that this is really so at least in the semiclassical
approximation. Namely if $H$ is written in the form $\exp\hbar^{-1}S$ where
$S=S_0+\hbar S_1+...$ we will find the asymptotics of $\int _L Hd\lambda$ as an
integral over some set of critical points of $S_0$ with the integrand expressed
in terms of Reidemeister torsion. This leads to a simplification of
quantization procedure and to the possibility to get rigorous results also in
the infinite-dimensional case, using the results of [4]. (We are talking about
the semiclassical approximation.)

 The present paper contains also a compete classification of $P$-manifolds and
$SP$-manifolds. The classification is interesting by itself, but in this paper
it plays also a role of an important tool in the proof of other results.
 \end{abstract}

  Let us consider a domain $U$ in a superspace $R^{n|n}$ with coordinates
$(x^1,...,x^n,\xi_1,...,\xi_n)$. An odd Poisson bracket (antibracket) of
functions $F$ and $G$ on $U$ can be defined by the formula
\begin {equation}
\{F,G\}={\partial_r F\over\partial x^a}\cdot{\partial_l G\over\partial\xi_a}-
{\partial_rF\over\partial\xi_a}\cdot {\partial_l G\over\partial x^a}
\end {equation}
where $\partial_r$ and $\partial_l$ denote the right derivative and the left
derivative correspondingly. (We suppose usually that $x^1,...,x^n$ are even and
$\xi_1,...,\xi_n$ are odd. However one can weaken this assumption by requiring
only that $x^a$ and $\xi_a$ have opposite parity.) The
 transformations of $U$ preserving the bracket (1) will be called $P$-
transformations (or odd symplectic transformations). Volume preserving
$P$-transformations (i.e. $P$-transformations having unimodular Jacobian) will
be called $SP$-transformations. (Of course we have in mind the supervolume.
Unimodularity of the Jacobian matrix means that the Berezinian of this matrix
is
equal to $1$.) $P$-manifold (or odd symplectic manifold) is by definition a
manifold pasted together from $(n|n)$-dimensional superdomains by means of
$P$-transformations. Replacing in this definition $P$-transformations by
$SP$-transformations we get the   definition of $SP$-manifold.\footnote {Using
the language of $G$-structures we can say that $P$-manifold is a supermanifold
provided with locally flat $P$-structure where $P$ is a group consisting of
linear transformations of $R^{n|n}$ preserving the bilinear form $x^i\xi_i$. To
get the definition of $SP$-manifold we have to replace here the group $P$ by
its
subgroup $SP=P\cap SL(n|n)$. We will not use the language of $G$-structures;
speaking about $P$-structure or $SP$-structure we will have in mind the
structure of $P$-manifold or $SP$-manifold.}
    In a general local coordinate system $(z^1,...,z^{2n})$ one can write the
Poisson bracket (1) in the form
\begin {equation}
\{F,G\}={\partial_r F\over \partial z^i}\omega^{ij}(z){\partial_l G\over
\partial z^j}
\end {equation}
where $\omega^{ij}(z)$ is an invertible matrix. Its inverse matrix
$\omega_{ij}(z)$ determines a differential form
\begin {equation}
\omega=dz^i\omega_{ij}dz^j
\end {equation}
 It is easy to check that this form is closed ($d\omega =0$). As in standard
symplectic geometry one can construct a vector field $K_H$ corresponding to a
function $H$ on a $P$-manifold $M$ by the formula
\begin {equation}
K_H^i(z)=\omega^{ij}(z){\partial_lH\over \partial z^j}
\end {equation}
By definition $K_H$ is a Hamiltonian vector field with Hamiltonian $H$. If the
function $H$ is odd then $K_H$ is even and vice versa.
The bracket (2) determines the structure of a Lie superalgebra on the linear
(super)space $F$ of (super)functions on $M$.
 It is easy to check that the map $H\rightarrow K_H$ is a homomorphism of $F$
into the Lie superalgebra $diff M$ of vector fields on $M$.
A submanifold $L\subset M$ is called isotropic if $\omega$ vanishes on $L$
(i.e.
$t^{\alpha}\omega_{ab}(x)\tilde {t}^b=0$ for every pair $t,\tilde{t}$ of
tangent
vectors to $L$ at the point $x\in L$). A Lagrangian manifold $L$ is by
definition an isotropic  manifold of dimension ($k|n-k$), $0\leq k\leq n$.

  One can give an invariant definition of $P$-manifold. Namely such a manifold
can be defined as an ($n|n$)-dimensional supermanifold provided with a \non\
$\omega$. This definition is equivalent to a previous one because one can prove
an analog of Darboux theorem: a \non\  $\omega$ locally can be written as
$dx^ad\xi _a$ by an appropriate choice of coordinates
$(x^1,...,x^n,\xi_1,...,\xi_n)-$  \dar. Moreover if $L$ is a \Lag\  of $M$ one
can choose \dar\  in the neighborhood of a point $a\in L$ in such a way that in
this neighborhood $L$ is singled out by the equations $x^{k+1}=...=x^n=0,\
\xi_1=...=\xi_k=0$. If we don't require that $x^i$ are even, $\xi_i$ are odd,
then we always can define a Lagrangian submanifold locally by the equations
$\xi_1=...=\xi_n=$0.

   The volume element in arbitrary coordinates ($z^1,...,z^{2n}$) can be
specified by means of the density function $\rho(z)$. In such a way
$SP$-structure on $M$ is determined by \non\ $\omega$ and by density $\rho$.
(It
is necessary to emphasize that $\rho$ is not arbitrary; one has to require that
in the neighborhood of every point in $M$ one can make $\rho\equiv 1$ by means
of appropriate choice of \dar.) As usual the volume element in $M$ determines
the divergence of vector field $K^a$ by the formula
\begin {equation}
div K=\rho ^{-1}{\partial _r(\rho K^a)\over
\partial z^a}={\partial_rK^a\over\partial z^a}+{\partial_r\ln \rho\over
\partial z^a}K^a
\end {equation}
Therefore one can define an operator $\Delta$ on the space $F$ of functions on
$SP$-manifold $M$ by the formula
\begin {equation}
\Delta H={1\over 2}div K_H
\end {equation}

  One can check that $\Delta^2=0$ using the existence of local coordinates with
$\omega=dx^ad\xi_a,\ \rho=1$. In these coordinates
\begin {equation}
\Delta={\partial_r\over\partial x^a}{\partial_l\over \partial \xi _a}
\end {equation}
and the relation $\Delta^2=0$ is evident. In a general coordinate system the
relation  $\Delta^2=0$ leads to conditions on $\rho$. One can prove that these
conditions are sufficient to assert that the \non\  $\omega$ and the density
function $\rho(z)$ determine an $SP$-structure; see Theorem 5 below. In the
formula (7) we suppose that the variables $x^a,\ \xi_a$ have opposite parity;
if
$x^1,...,x^n$ are even and $\xi_1,...,\xi_n$ are odd as we assume usually
the right derivative with respect to $x^a$ in (7) is of course the standard
derivative.
If $L$ is a \Lag\  of $SP$-manifold $M$ one can define a volume element in $L$
(up to a sign). Namely, if in \dar\
 $x^1,...,x^n, \xi_1,...,\xi_n$ the manifold  $L$ is singled out by the
equations $x^{k+1}=...=x^n=0,  \ \xi_1=...=\xi_k=0$ then the volume element in
$L$ can be defined as $dx^1...dx^kd\xi_{k+1}...d\xi_n$. It is easy to check
that
this volume element is well defined up to a sign. (This fact follows
immediately
from another description of the volume element in $L$ given in the proof of the
Lemma 4.) One has to impose some global conditions to define the volume element
globally. Namely, we will prove that it suffices to require that $M$ be
orientable, i.e. that  $m(M)$ be orientable. We denote by $m(M)$ the body of
$M$
, i.e. the bosonic part of $M$.

  The Batalin-Vilkovisky approach to quantization is based on the following
theorem: if $L_0$ and $L_1$ are closed oriented \Lag s connected with a smooth
family of closed oriented \Lag s $L_t$ and an even function $H$ on $M$
satisfies
the condition $\Delta H=0$ then $\int _{L_0}Hd\lambda_0=\int_{L_1}Hd\lambda_1$.
  For completeness we will sketch a proof of this statement. As usual it is
sufficient to consider an infinitesimal deformation of the Lagrangian manifold
$L$; moreover one can assume that $L$ is deformed only in a domain where (after
appropriate change of coordinates) it is singled out by equations
$\xi_1=...=\xi_n=0$ and where $\rho=1$. Then the deformed manifold can be
specified by means of an odd function $\Psi (x^1,...,x^n)$ that vanishes
outside
of this domain. Namely, the deformed manifold can be defined by the equations
$\xi_j={\partial_l\Psi\over\partial x^j}$. The variation of the integral
$\int_L
Hd\lambda$ by this deformation can be written as
  $$ \int {\partial_rH\over\partial x^j}
  {\partial_l\Psi\over\partial\xi_j}dx^1...dx^n.$$

  Integrating by parts and using $\Delta H=0$ we obtain that this variation is
equal to zero.

  In the formulation of Batalin-Vilkovisky theorem we assume that the volume
elements $d\lambda_0$ and $d\lambda_1$ in $L_0$ are choosen in an appropriate
way; namely we require the existence of volume elements $d\lambda_t$ in $L_t$
depending continuously on $t$ and connecting $d\lambda_0$ and $d\lambda_1$. A
similar assumption must be made about the orientation of $L_0$ and $L_1$. Our
aim is to prove a generalization of this theorem. Namely, we will prove the
following:

  {\bf Theorem 1.} Let $L_0$ and $L_1$ be closed oriented \Lag s of an
orientable $SP$-manifold $M$. If the cycles $m(L_0)$ and $m(L_1)$ are
homologous
in $m(M)$ over $R$ (i.e. $m(L_0)$ and $m(L_1)$ determine the same element of
$H_k(m(M),R)$) then
\begin {equation}
\int_{L_0}Hd\lambda_0=\int_{L_1}Hd\lambda_1
\end {equation}
for every function $H$ satisfying $\Delta H=0$.

  We will prove also

  {\bf Theorem 2.} If $H=\Delta K$ then for every closed Lagrangian manifold
$L$
\begin {equation}
\int_L Hd\lambda=0.
\end {equation}

  The proof of these theorems will be based on an explicit description of
$P$-manifolds and their \Lag s. I don't know any direct proof of these
theorems.

  We begin with the remark that every transformation $\tilde {x}=f(x)$ of
$n$-dimensional domain with coordinates $x^1,...,x^n$ can be extended to a
$P$-transformation  $(x^1,...,x^n,\xi _1,...,\xi_n)\rightarrow(\tilde
{x}^1,...,\tilde{x}^n,\tilde{\xi}_1,...,\tilde{\xi}_n)$ by means of the formula

\begin {equation}
\tilde{\xi}_i={\partial x^j\over \partial\tilde {x}^i}\xi_j.
\end {equation}
This means that a cotangent bundle $T^*N$ to an $n$-dimensional manifold $N$
has
a natural structure of $P$-manifold (the formula (10) coincides with
transformation low of covectors).
We will prove

  {\bf Theorem 3.} Every $(n|n)$-dimensional $P$-manifold $M$ is equivalent to
a $P$-
manifold of the form $T^*N$. Namely,one can take $N=m(M)$.

  Let us begin with a remark that for every $m$-dimensional vector bundle
$\alpha$ over an $n$-dimensional manifold $N$ one can construct an
$(m|n)$-dimensional supermanifold considering the fibres as odd linear spaces.
More precisely,if a vector bundle over $N$ has transition functions $\tilde
{x}^i=f^i(x^1,...,x^n)$, $\tilde{\eta}^l=\alpha_k^l(x^1,...,x^n)\eta^k$, where
$x^i$ are coordinates in the base, $\eta^k$ are coordinates in the fibre, one
can construct a  supermanifold pasted together by means of the same formulas
where $\eta^k$ are considered as odd coordinates. It is well known that every
real $m|n$-dimensional supermanifold can be obtained by means of this
construction [5]; therefore we can assume that $P$-manifold $M$ is a bundle
$\alpha$ over $N=m(M)$. (The bundle $\alpha$ has an invariant description as so
called conormal bundle [5], we will not use this description.) Sometimes we
will
use the notation $N_{\alpha}$ for the supermanifold corresponding to the bundle
$\alpha$ over $N$.
 Let us restrict the form $\omega$ specifying the $P$-structure in
$M=N_{\alpha}$ to $N\subset M$ (i.e. we take $\eta =0$). The expression
\begin {equation}
\omega|_{\eta=0}=\omega_{ij}(x)dx^id\eta^j
\end {equation}
determines a non-degenerate pairing between fibres of $\alpha$ and tangent
spaces to $N$. The existence of this pairing permits us to identify $\alpha$
with cotangent bundle and $M$ with $T^*N$. However it is possible a priori that
the $P$-structure on $T^*N$ arising from this identification and the standard
$P$-structure on $T^*N$ are different. To show the equivalence of these
$P$-structures we note that corresponding forms $\omega$ and $\omega_0$ can be
connected by a smooth family $\omega_t=(1-t)\omega_0+t\omega$ of closed
non-degenerate odd forms. (To check that the forms $\omega_t$ are
non-degenerate we use the fact that $\omega$ and $\omega_0$ coincide on $N$
imbedded in standard way into $T^*N$.  Non-degeneracy of $\omega_t$ on $T^*N$
follows from  non-degeneracy on $N\subset T^*N$.) To finish the proof we
 utilize the following:

  {\bf Lemma 1.} If $\omega$ is a \non\  and $\sigma$ is a closed odd 2-form
then one can find a vector field $V$ in such a way that $\sigma=L_V\omega$
where
$L_V\omega$ is the Lie derivative of $\omega$ with respect to $\omega$ (the
change of $\omega$ by the infinitesimal transformation $V$).

 To prove this lemma we note that $L_V\omega$ can be represented as
\begin {equation}
L_V\omega=(d\omega)_V-d\omega_V
\end {equation}
where for every $k$-form $\sigma$ we denote by $\sigma_V$ the $(k-1)$-form
obtained from $\sigma$ by contraction with the vector field $V$. For example,
if
$\omega=dz^i\omega_{ij}dz^j$ then $\omega_V=V^i\omega_{ij}dz^j$. If $\omega$ is
closed then $L_V\omega=-d\omega_V$. Every closed odd 2-form $\sigma$ is exact:
$\sigma=d\lambda$, where $\lambda=\lambda_jdz^j$. It remains to say that $V^i$
can be found from the equation
\begin {equation}
\lambda_j=-V^i\omega_{ij}
\end {equation}
This equation always has a solution because $\omega_{ij}$ is
non-degenerate. Moreover this solution is unique.

 The proof of the lemma repeats the standard proof of the fact that an even
symplectic structure on closed manifold does not change if the $2$-form
defining
it changes, but corresponding cohomology class remains intact.

  If $\omega_t$ is a smooth family of \non s on $M=T^*N$ coinciding on
$N\subset
M$ it follows immediately from the lemma that all these forms determine
equivalent $P$-structures. The lemma shows that an infinitesimal variation of
form $\omega $ gives an equivalent $P$-structure. The study of a smooth
deformation of $\omega$ can be reduced to the study of infinitesimal variation.
To construct the transformations proving the equivalence we have to solve
differential equation
\begin {equation}
\dot {z}(t)=V(t)z(t)
\end {equation}
where the vector field $V(t)$ satisfies
\begin {equation}
\dot {\omega}_t=L_{V(t)}\omega_t
\end {equation}
It follows from the proof of the lemma that one can find $V(t)$ is such a way
that it will be differentiable with respect to $t$. This assumption  guarantees
the existence of solution to (14). (In the case of even symplectic structure it
is necessary to assume compactness of symplectic manifold to guarantee the
existence of solution to the analog of (14). In the case at hand we don't need
this  assumption because $V^i$ generates a zero vector field on the body $N$ of
$M=T^*N$.)

   In what follows we restrict ourselves by the case when the $P$-manifold $M$
is realized as $T^*N$ with standard $P$-structure; as we proved this can be
made
without loss of generality. Let us define standard  \Lag s of $T^*N$ in the
following way. Let us suppose that $K$ is a $k$-dimensional submanifold of $N$.
Then we can construct an $(n-k)$-dimensional bundle $\lambda$ over $K$
consisting of covectors orthogonal to $K$. Supermanifold $L_K$ corresponding to
this bundle is naturally imbedded into $T^*N$ and can be considered as
$(k|n-k)$-dimensional \Lag\ of $T^*N$.

  {\bf Theorem 4.} For every \Lag\ of $T^*N$ one can find a smooth deformation
of this submanifold into a standard \Lag \ (i.e. into a submanifold of the form
$L_K$)

  To prove this theorem we consider at first the group $G_M$ of all
transformations of arbitrary supermanifold $M$. Without loss of generality we
assume that $M=N_{\sigma}$ where $\sigma$ is a vector bundle over a manifold
$N$. Let us denote by $G_{\sigma}$ the group of automorphisms of the bundle
$\sigma$. In local coordinates these automorphisms are given by formulas
$\tilde
{x}^i=F^i(x^1,...,x^n),\ \tilde {\eta}^j=a^j_i(x^1,...,x^n)\eta^i$, where
$a^j_i$ is a non-degenerate matrix. The same formulas determine
transformations of a supermanifold $N_{\sigma}$; therefore we have a natural
imbedding $i$ of $G_{\sigma}$ into $G_M$. There exists also a natural map $\pi$
of $G_M$ onto $G_{\sigma}$.  In local coordinates a transformation of $M$ can
be
written as
\begin {equation}
\tilde{x}^i=f^i(x^1,...,x^n)+\sum_{k=1}\sum_{j_1,...,j_{2k}}f^i_{j_1,...,j_{2k}}(x^1,...,x^n)\eta^{j_1}...\eta^{j_{2k}},
\end {equation}
\begin {equation}
\tilde{\eta}^j=a^j_i(x^1,...,x^n)\eta^i+\sum_{k=1}\sum_{i_1,...,i_{2k+1}}a^j_{i_1,...,i_{2k+1}}(x^1,...,x^n)\eta^{i_1}...\eta^{i_{2k+1}}.
\end {equation}

  Leaving only the first term in (16),(17) we get an automorphism of $\sigma$.
(In more invariant words one can say that the transformation of a supermanifold
generates naturally an automorphism of corresponding conormal bundle). It is
easy check that the maps $i$ of $G_{\sigma}$ into $G_M$ and $\pi$ of $G_M$ onto
$G_{\sigma}$ generate a homotopy equivalence between $G_M$ and $G_{\sigma}$.
The
main fact leading to this conclusion is that (16),(17) determine a
transformation of $M$ by any choice $f^i_{j_1,...,j_2k},a^j_{i_1,...,i_{2k+1}}$
for $k\geq 1$ if $f^i(x^1,...,x^n)$ and $a^j_i(x^1,...,x^n)$ determine  an
automorphism of $\sigma$. Therefore we can simply multiply all these functions
by $\tau,\ 0\leq \tau\leq 1$, to obtain a family $Q_{\tau},\ 0\leq \tau \leq
1$,
of  transformations of $M$ obeying $Q_1=i\alpha,\ Q_0=i\pi,\ \pi Q_{\tau}=\pi$.

  If $M$ is a $P$-manifold, we will denote by $S_M$ the group of all
$P$-transformations of $M$ (transformations preserving the $P$-structure in
$M$). In this case $\sigma$ is a cotangent bundle and $G_{\sigma}$ is imbedded
in $S_M$. One can prove the following lemma which is interesting by itself.

   {\bf Lemma 2.} The imbedding $i$ of  $G_{\sigma}$ into $S_M$ and the natural
map $\pi$ of $S_M$ onto  $G_{\sigma}$  determine a homotopy equivalence between
$G_{\sigma}$ and $S_M$.

    To prove this statement we use the deformation $Q_{\tau}$ constructed above
and the arguments used in the proof of Lemma 1. Namely, we will define the
deformation $\tilde {Q}_{\tau}$ as $R_{\tau} Q_{\tau}$ where $R_{\tau}$ is a
transformation of $M=T^*N$ satisfying $(R_{\tau}Q_{\tau})^*\omega=\omega$. Such
a transformation $R_{\tau}$ can be found by solving the equations (14),(15). To
guarantee the continuity of $R_{\tau}$ with respect to $\tau$ we have to
eliminate the freedom in the construction of $R_{\tau}$. This can be made if we
construct  $R_{\tau}$ by means of (15) with $V_t$ found as a solution of the
equation $\lambda_j^t=-V^i_t\cdot\omega_{ij}$ where $\lambda
^t=\lambda^t_jdz^j$
is specified by the formula $\lambda^t=(Q^*_t)^{-1}\lambda,\ d \lambda=\omega$.

    Now we are able to prove the Theorem 4. It is easy to check that for
arbitrary \Lag\ $L$ of $P$-manifold $M$ one can find a map $\varphi$ of $T^*L$
into $M$ preserving $P$-structure. Representing $L$ as $K_{\beta}$ where
$K=m(L)$, $\beta$ is a vector bundle over $K$ we can construct a map of $L$
onto
a standard \Lag\ of $M=T^*N$ and extend this map to a map $\tilde{\varphi}$ of
$T^*L$ into $M$ preserving $P$-structure. Using the Lemma 2 we can deform
$\varphi$ into $\tilde {\varphi}$ and therefore every \Lag\ into a standard
\Lag.

     Let us consider a manifold $N$ provided with a volume element $\alpha$
(one
can consider $\alpha$ as an $n$-form $\alpha(x^1,...,x^n)dx^1\wedge ...\wedge
dx^n$ where $\alpha(x^1,...,x^n)$ is a non-vanishing function on $N$).
Differential forms on $N$ can be considered as functions on a supermanifold
$TN$, corresponding to the tangent bundle over $N$. If $x^1,...,x^n$ are local
coordinates in $N$ then  $x^1,...,x^n,\ \eta^1=dx^1,...,\eta^n=dx^n$ can be
considered as coordinates in $TN$. If $\omega$ is a differential form on $N$ (a
function on $TN$) one can define a function $\tilde{\omega}=F\omega$ on $T^*N$
by the formula
\begin {equation}
\tilde{\omega}(x,\xi)=\int e^{\xi_i\eta^i}\omega(x,\eta)\alpha(x)d^n\eta
\end {equation}
In other words $\tilde{\omega}$ is a Fourier transform of $\omega$ with respect
to odd variables. (The functions on $T^*N$ can be identified with antisymmetric
polyvector fields. In this interpretation Fourier transformation is simply the
duality transformation, i.e. contraction of covector field $\omega_{i_1...i_k}$
with the universal antisymmetric tensor $\alpha\epsilon^{j_1...j_n}$.) It is
easy to check that
\begin {equation}
{\partial_l (F\omega)\over\partial\xi_i}=F(\eta^i\omega),\ \ {\partial
(F\omega)\over\partial x^i}=F({\partial\omega\over\partial
x_i})+\alpha^{-1}{\partial\alpha\over\partial x_i}F(\omega).
\end {equation}
Using these formulas we obtain that
\begin {equation}
F(d\omega)=\Delta F(\omega)
\end {equation}
where $d$ denotes the exterior differential of $\omega$ (in the language of
functions on $TN$ we have $d=\eta^i{\partial\over\partial x^i}$) and the
operator $\Delta$ is constructed by means of $SP$-structure on $P$-manifold
$T^*N$, specified by the volume element
\begin {equation}
\rho_0(x,\xi)d^nxd^n\xi =\alpha^{-2}(x)d^nxd^n\xi.
\end {equation}
(This connection between $d$ and $\Delta$ was used in [3]).

   Now we able to prove Theorems 1 and 2 for the case when the manifold
$M=T^*N$
is provided with standard $P$-structure and with the volume element (21). We
will use the following statement that can be easily proved in this case.

  {\bf Lemma 3.} If $\omega$ is a form on $N$ and $K$ is a closed oriented
submanifold
of $N$ then
\begin {equation}
\int_K\omega=\int_{L_K}F(\omega)d\lambda
\end {equation}
where $L_K$ denotes the \Lag \ of $T^*N$ corresponding to $K$.

    The proof of the Lemma 3 in the case when $\omega$ has a support in a
domain
where $K$ in an appropriate coordinate system can be singled out by equations
$x^{k+1}=0,...,x^n=0$ is immediate. Without loss of generality one can assume
that $\omega$ is a monomial with respect to $\eta^1,...,\eta^n$. Only the
monomial $\omega=\gamma(x)\eta^1...\eta^k$ gives a non-zero contribution to the
integrals in (22). For this monomial we have
$$F(\omega)=\gamma(x)\alpha(x)\xi_{k+1}...\xi_n.$$
The volume element $d\lambda$ on $L_K$ induced by (21) can be written in the
form
$$d\lambda=\alpha(x)^{-1}dx^1...dx^kd\xi_{k+1}...d\xi_n$$
(we omit the proof of this assertion because a more general fact will be proven
later; see Lemma 4 ). Using the expressions for $F(\omega)$ and $d\lambda$ we
obtain (22) in the case at hand. The general case can be reduced to this
simplest case by means of standard technique (one has to use the partition of
unity).

 The statements of Theorems 1 and 2 follow immediately from (22) and (20) when
the \Lag s are standard. The case of general \Lag s of $T^*N$ can be reduced to
this simplest case by means of Theorem 3. Therefore we can say that Theorems 1
and 2 are proved in the case when $SP$-structure in $T^*N$ is determined by the
density that does not depend on $\xi$. (The volume element corresponding to
such
a density can be represented up to a sign in the form (21).)

    In the consideration above we did not pay sufficient attention to the
choice
of the sign of the volume element in \Lag . It suffices to analyze this
question
for standard \Lag\ $L_K$. Let us introduce the notation $\Lambda(E)$ for the
one-dimensional linear space of translationally invariant real measures in the
linear superspace $E$. In other words $\Lambda(E)$ consists of functions of
bases in $E$ having degree $1$ (i.e. to specify an element
$\alpha\in\Lambda(E)$
we have to assign to every basis $e\in E$ a real number $\alpha(e)$ in such a
way that $\alpha(Ae)=\det A\cdot\alpha(e)$ where $\tilde {e}=Ae$ denote a basis
obtained from $e$ by means of a linear transformation: $\tilde{e}_i=A^j_ie_j$.)
To specify the volume element in $L_K$ one has to single out a non-zero element
of $\Lambda(TL_K(z))-$ a non-zero measure in tangent space $TL_K(z)$ to $L_K$
at
every point $z\in L_K$. One can identify $\Lambda(TL_K(z))$ with
$\Lambda(TK(m(z)))\otimes\Pi \Lambda(TN(m(z))/TK(m(z)))^*=\Lambda(TN(m(z))$.
(Here $\Pi$ denotes the parity reversion. We used that $\Lambda(\Pi
E)=\Lambda(E)^*,\ \Lambda(E^*)=\Lambda(E)^*,\
\Lambda(E_1)=\Lambda(E_2)\otimes\Lambda(E_1/E_2)$ if $E_2\subset E_1$). The
spaces $\Lambda(TL_K(z))$ can be considered as fibres of a line bundle over
$L_K$. If this bundle is trivial and locally the volume element is defined up
to
a sign then the volume element can be defined globally. Conversely if the
volume
element is defined globally it can be considered as a non-zero section of this
bundle and this bundle is trivial. Using the identification
 $\Lambda TL_K(z)=\Lambda TL_K(m(z))=\Lambda(TN(m(z)))$ we conclude that in the
case when $N$ is orientable (i.e. the bundle over $N$ with the fibres
$\Lambda(TN(x))$ is trivial) the volume element on every \Lag \ can be defined
globally. In the general case \Lag \ $L$ of $SP$-manifold $M$ can be provided
with global volume element if and only if its body $m(L)$ can be imbedded in an
orientable submanifold of $m(M)$.

 Now we have to give a proof of Theorems 1 and 2  for general $SP$-manifold.
The
proof is based on the following

   {\bf Lemma 4.} Let us suppose that  $SP$-structure in a  $SP$-manifold $M$
is
specified by the density $\rho$. If the density $\tilde{\rho}=\rho e^{\sigma}$
in $M$ also determines a $SP$-structure in $M$ then
\begin {equation}
\Delta_{\rho}\sigma+{1\over 4}\{\sigma,\sigma\}=0
\end {equation}
where $\Delta_{\rho}$ denotes the operator $\Delta$ corresponding to the
$SP$-structure determined by the density $\rho$. The operator $\Delta$
corresponding  to the density $\tilde{\rho}$ can be written in the form
\begin {equation}
\Delta_{\tilde{\rho}}=e^{-\sigma/2}\Delta_{\rho}(e^{\sigma/2}H).
\end {equation}
Volume elements $d\tilde{\lambda}$ and $d\lambda$ in the \Lag \ $L\subset M$
corresponding to  $SP$-structures at hand are connected by the formula
\begin {equation}
d\tilde{\lambda}=e^{\sigma/2}d\lambda.
\end {equation}

  The statement of the Lemma 4 is local and therefore we can simplify the proof
using \dar \ and assuming that $\rho=1$. We can write in these coordinates
\begin {equation}
\Delta_{\tilde{\rho}}H=\Delta_{\rho}H+{1\over
2}\{\sigma,H\}={\partial\over\partial x^a}{\partial_e\over\partial \xi
^a}H+{1\over 2}\{\sigma,H\}.
\end {equation}
Calculating $\Delta_{\tilde{\rho}}^2 $ we get
\begin {equation}
\Delta_{\tilde{\rho}}^2H=\{\Delta_{\rho}\sigma+{1\over
4}\{\sigma,\sigma\},H\}=0
\end {equation}
 This equation shows that in the case when the density $\tilde{\rho}$
determines
an  $SP$-structure, $\Delta_{\rho}\sigma+\{\sigma,\sigma\}/4={\rm const} $.
This equation can be written also in the form
 \begin {equation}
\Delta_{\rho}e^{\sigma/2}={\rm const}\cdot e^{\sigma/2}.
\end {equation}

  Applying $\Delta_{\rho}$ to (28) we obtain from $\Delta^2_{\rho}=0$ that the
constant in this equation is equal to $0$. In such a way

  \begin {equation}
  \Delta _{\rho}e^{\sigma/2}=0.
  \end {equation}
Using (29) one can check that (24) follows from (23).
  To prove (25) we will give another description of volume element in \Lag\
$L$.
Let us fix a basis $(e_1,...,e_n)$ in the tangent space $TL(z)$ to $L$ at the
point $z\in L$. Then one can find a basis $(e_1,...,e_n,f^1,...,f^n)$ in the
tangent space $TM(z)$ to $M$ satisfying $\omega(e_i,f^j)=\delta_i^j$
($P$-structure  in $M$ determines an odd bilinear form $\omega$ on $TM(z)$).
The
volume element $\lambda$ in $L$ can be defined by the formula
\begin {equation}
\lambda(e_1,...,e_n)=\mu(e_1,...,e_n,f^1,...,f^n)^{1/2}
\end {equation}
where $\mu$ denotes the volume element determined by $SP$-structure in $M$. The
equation (25) follows immediately from (30).

   Using the Lemma 4 we can reduce the study of
$SP$-structure with density function $\tilde{\rho}=\rho e^{\sigma}$ to the
study
of $SP$-structure with density function $\rho$. In particular we are able now
to
prove the Theorems 1 and 2 for all $SP$-manifolds. As we mentioned already it
is
sufficient to consider manifolds of the form $M=T^*N$ with standard
$P$-structure. If $SP$-structure in $M$ is specified by the density
\begin {equation}
\rho(x,\xi)=\rho_0(x)+\sum_{k>1}\rho ^{i_1...i_k}(x) \xi_{i_1}...\xi_{i_k}
\end {equation}
we can consider another $SP$-structure in $M$ determined by the density
$\rho_0(x)$. It is clear that $\rho(x,\xi)=\rho_0(x)e^{\sigma(x,\xi)}$ where
$\sigma(x,\xi)=0$ for $\xi_1=...=\xi_n=0$. It follows from (25) that for every
\Lag \ $L\subset M$ we have
\begin {equation}
\int_LHd\lambda=\int_LHe^{\sigma(x,\xi)/2}d\lambda_0.
\end {equation}
Further if $\Delta H=0$ we obtain from (24) that $\Delta_0(He^{\sigma/2})=0$
and
if $H=\Delta K$ we get that $He^{\sigma/2}=\Delta_0(Ke^{\sigma /2})$. (We use
the notations $d\lambda$ and $d\lambda_0$ for volume elements in $L$ determined
by the densities $\rho$ and $\rho_0$; the notations $\Delta$ and $\Delta_0$
have
similar meaning.) Using these remarks we reduce the proof of Theorems 1 and 2
for the density  $\rho$ to the case of density $\rho_0$. This case was analyzed
already.

  The consideration above permits us to construct one-to-one correspondence
between $SP$-structures in connected $P$-manifold $M$ and cohomology classes
$s\in H(m(M),R)$ satisfying $s^n\not= 0$. (Recall that by definition $H(N,R)$
is
the direct sum of $k$-dimensional cohomology groups $H^k(N,R)$; we represent
$s\in H(m(M),R)$ as $s^0+s^1+...+s^n$ where $s^k\in H^k(m(M),R).$) We suppose
without loss of generality that $M$ coincides with $T^*N$ provided with
standard
$P$-structure. Let us fix a volume element $\alpha$ in $N$. If $\omega$ is a
differential form $\omega =\sum_{k=0}^n\omega^k$ in $N$ (i.e. a function
$\omega(x,\eta )$ on $TN$) we define a function $\tilde{\omega}(x,\xi)$ on
$T^*N$ by means of Fourier transformation (18). Let us assume that the
$n$-dimensional component $\omega^n$ of the form $\omega$ does not vanish (i.e.
$\omega^n=\beta(x)dx^1\wedge...\wedge dx^n$ where $\beta(x)\not= 0$). We will
define the density function $\rho_{\omega}(x,\xi)$ on $T^*N$ by the formula
\begin {equation}
\rho_{\omega}(x,\xi)=\alpha^{-2}(x)\tilde{\omega}(x,\xi)^2.
\end {equation}
It is easy to check that
$\rho_{\omega}(x,\xi)$ does not depend on the choice of $\alpha$:
\begin {equation}
\rho_{\omega}(x,\xi)=(\int e^{\xi_i\eta^i}\omega(x,\eta)d^n\eta)^2.
\end {equation}

 {\bf Theorem 5.} The density function
$\rho_{\omega}(x,\xi)=\alpha^{-2}(x)\tilde{\omega}(x,\xi)^2$ determines an
 $SP$-structure in $T^*N$ if and only if the form $\omega$ is closed and its
$n$-dimensional component $\omega^n$ does not vanish. Every $SP$-structure in
$T^*N$ can be described by means of density function of such a kind. If
$\rho_{\omega}$ and $\rho_{\omega^{\prime}}$ are density functions
corresponding  to closed forms $\omega$ and $\omega^{\prime}$ then
corresponding
 $SP$-structures are equivalent only in the case when the form
$\omega^{\prime}-\omega$ is exact.

  We say here that two $SP$-structures on $T^*N$ are equivalent if there exists
a $P$-transformation connecting these $SP$-structures and homotopic to the
identity mapping. (As we have seen the transformation of the supermanifold
$T^*N$ is homotopic to identity if and only if corresponding transformation of
$N$ is homotopic to identity.)

  Let us begin the proof with the remark that the application of Lemma 4 to
$\tilde{\rho}=\rho_{\omega},\ \rho_0=\alpha^{-2},\
\tilde{\omega}(x,\xi)=e^{\sigma/2}$ shows that the operator $\Delta$
corresponding to the density $\rho_{\omega}$ satisfies $\Delta^2=0$ if and only
if the form $\omega$ is closed. Therefore if $\rho _{\omega}$ determines an
$SP$-structure then $\omega$ is closed. To prove that in the case of closed
$\omega$
 the density $\rho_{\omega}$  determines an  $SP$-structure we will construct a
family $\omega_t$ of closed forms: $\omega_t=(1-t)\omega^n+t\omega$,
corresponding densities $\rho_t=\rho_{\omega_t}$ and operators $\Delta_t$
defined by the formula
 \begin {equation}
 \Delta_tH=e^{-\sigma_t/2}\Delta_0(e^{\sigma_t/2}H)=\tilde{\omega}^{-1}_t\cdot
\Delta_0(\tilde{\omega}_tH).
\end {equation}
 Here  $\Delta_0$ is constructed by means of
  the density $\rho_0$ corresponding to the form $\omega_t|_{t=0}=\omega^n$. It
is clear that the density $\rho_0$ determines an  $SP$-structure. To prove that
$\rho_{\omega}$ also determines an $SP$-structure it is sufficient to check
that
at least locally we can transform $\rho_{\omega}$ into $\rho_0$ by means of
$P$-transformation. To find such a $P$-transformation we construct at first an
infinitesimal $P$-transformation (Hamiltonian vector field) transforming
$\rho_t$ into $\rho_{t+dt}$. To verify the existence of such a field we note
that the change of density $\rho_t$ by the infinitesimal transformation
generated by the vector field $K_t$ can be written as
\begin {equation}
{\partial_r\over\partial z^a}(\rho_tK^a_t)=\rho_t{\rm div} K_t
\end {equation}
If $K_t$ is a Hamiltonian vector field with Hamiltonian $H_t$ we obtain
\begin {equation}
\dot {\rho}_t=2\rho_t\Delta_tH_t
\end {equation}
or equivalently
\begin {equation}
\dot {\sigma}_t=2\Delta_tH_t=2e^{-\sigma_t/2}\Delta_0(e^{\sigma_t/2}H_t).
\end {equation}
{}From the other side we obtain from Lemma 4 that $\Delta_0e^{\sigma_t/2}=0$
Differentiating this equation with respect to $t$ we get
\begin {equation}
\Delta_0(\dot {\sigma}_te^{\sigma_t/2})=0
\end {equation}
It follows from (39) that (38) considered as an equation for $H_t$ can be
solved
at least locally. (One should make the Fourier
transformation (18) and use the Poincare lemma). As usual to find the
transformation connecting
$\rho_{\omega}$ and $\rho_0$ we have to integrate the equation
$\dot{z}=K_t(z)$.

In such a way we proved that the density (33) determines an $SP$-structure if
$\omega$ is closed. The same arguments can be used to check that the
$\rho_{\omega^{\prime}}$ and $\rho_{\omega}$ determine equivalent
$SP$-structures if $\omega^{\prime}-\omega$ is exact (it follows from the
exactness of $\omega^{\prime}-\omega$ that the equation (38) for $H_t$ can be
solved globally). To finish the proof of Theorem 5 we have to check that in the
case when $\omega^{\prime}-\omega$ is not exact the densities
$\rho_{\omega^{\prime}}$ and $\rho_{\omega}$ cannot determine equivalent
$SP$-structures. Let us suppose that there exists a  $P$-transformation $Q$
connecting $\rho_{\omega^{\prime}}$ and $\rho_{\omega}$ and homotopic to
identity. One can conclude from Lemma 2 that in this case we can find a smooth
family $Q_t$
of  $P$-transformations connecting $Q=Q_1$ with the identity map $Q_0$. Let us
denote by $\rho_t$ the density obtained from $\rho_{\omega}$ by means of $Q_t$;
corresponding form will be denote by $\omega_t$. The density $\rho_{t+dt}$ can
be obtained from the density $\rho_t$ by means of infinitesimal
$P$-transformation (Hamiltonian vector field $K_t=\dot{Q}_tQ^{-1}_t$) and we
can
apply (38). It follows from (38) that the form $\dot {\omega}_t$ is
exact,therefore the form $\omega^{\prime}-\omega=\int_0^1\dot{\omega}_tdt$ is
exact too.

  I am indebted to A.Givental,M.Kontsevich, A.Weinstein and E.Witten for useful
discussions.
  \vskip .1in
  \centerline{{\bf References}}
  \vskip .1in
  1. Batalin, I., Vilkovisky, G.: Gauge algebra and quantization. Physics
Letters, 102B,27(1981)

  2. Batalin, I., Vilkovisky, G.: Quantization of gauge theories with linearly
dependent generators. Phys. Rev. D29,2567(1983)

  3. Witten, E.: A note on the antibracket formalism. Preprint IASSHS-HEP-9019

  4. Schwarz, A.: The partition function of a degenerate functional. Commun.
Math. Phys. 67,1 (1979)

  5.Berezin, F.:Introduction to algebra and analysis with anticommuting
variables, Moscow Univ.,1983 (English translation is published by Reidel).
\end{document}